# Secure communication between UAVs using a method based on smart agents in Unmanned Aerial Vehicles

Maryam Faraji-Biregani[1] . Reza Fotohi[2]

**Abstract** Unmanned Aerial Vehicles (UAVs) can be deployed to monitor very large areas without the need for network infrastructure. UAVs communicate with each other during flight and exchange information with each other. However, such communication poses security challenges due to its dynamic topology. To solve these challenges, the proposed method uses two phases to counter malicious UAV attacks. In the first phase, we applied a number of rules and principles to detect malicious UAVs. In this phase, we try to identify and remove malicious UAVs according to the behavior of UAVs in the network in order to prevent sending fake information to the investigating UAVs. In the second phase, a mobile agent based on a three-step negotiation process is used to eliminate malicious UAVs. In this way, we use mobile agents to inform our normal neighbor UAVs so that they do not listen to the data generated by the malicious UAVs. Therefore, the mobile agent of each UAV uses reliable neighbors through a three-step negotiation process so that they do not listen to the traffic generated by the malicious UAVs. The NS-3 simulator was used to demonstrate the efficiency of the SAUAV method. The proposed method is more efficient than CST-UAS, CS-AVN, HVCR, and BSUM-based methods in detection rate, false positive rate, false negative rate, packet delivery rate, and residual energy.

**Keywords** Unmanned Aerial Vehicles (UAVs) . Malicious UAVs. Routing security . Three-step negotiation process

✉ Maryam Faraji-Biregani
 m.faraji@ashrafi.ac.ir

✉ Reza Fotohi*
 R_fotohi@sbu.ac.ir; Fotohi.reza@gmail.com

[1] Department of Computer Engineering, Faculty of Engineering, Shahid Ashrafi Esfahani University, Isfahan, Iran
[2] Faculty of Computer Science and Engineering, Shahid Beheshti University, G. C. Evin, Tehran, Iran

# 1 Introduction

Today, UAVs have made significant progress in military and defense areas such as reconnaissance, surveillance, and security missions as well as in civilian areas such as urban planning, search, law enforcement, traffic monitoring, accident management, agricultural assessment, and entertainment. Environmental monitoring Photography, infrastructure monitoring and rescue operations are growing rapidly [1-4]. Because UAVs have security vulnerabilities, hackers exploit this security vulnerability. A UAV that is infiltrated will have a negative performance because it will be controlled by intruders. Also, UAVs, despite their many advantages, have been less studied to solve the security problem, so they can be easily penetrated by enemies. Therefore, with UAV hacking, consequences such as loss of top-secret information, damage to infrastructure, and loss of essential missions become apparent. Most intrusions and threats are usually due to weak security in communication protocols.

In this paper, the following two phases based on the hash function are used to secure the communication between the UAVs against blackhole, sinkhole and gray hole attacks. In the first phase, in order to prevent the sending of fake UAV information to other normal UAVs, a series of safety rules were used to identify malicious operations to identify malicious UAVs and exclude them from routing operations. In the second phase, a mobile agent based on a three-step negotiation process was used to prevent the normal UAV from receiving false information from other malicious UAVs. The combination of these two phases led to secure communications between the UAVs so that packets could exchange information securely. The hash function is also used to detect fake agents and interact with authentic UAVs in the network. Figure 1 shows an example of communication between UAVs that a malicious UAV can exploit the information exchanged between other UAVs and produce fake information on the network.

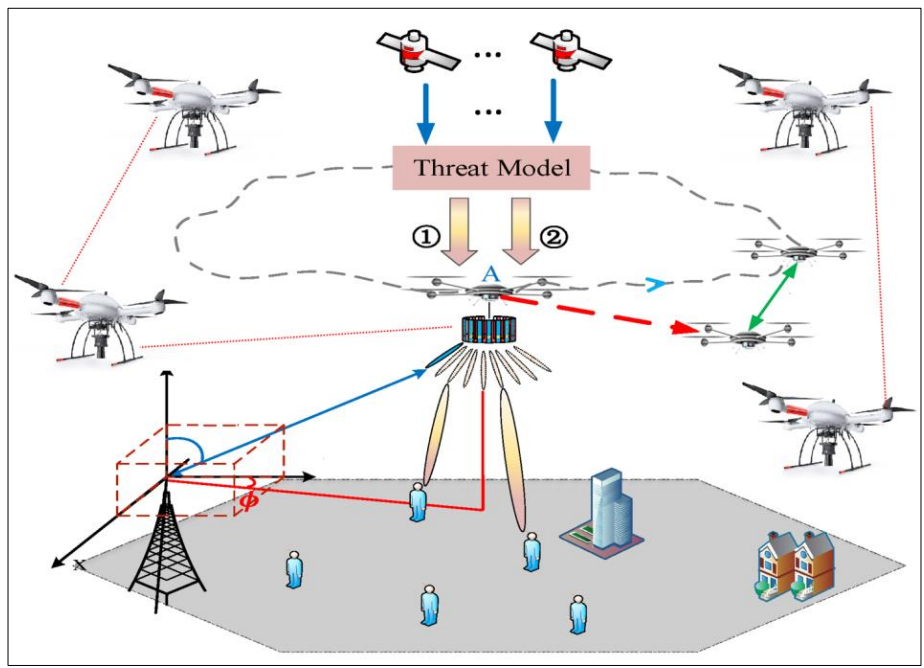

**Fig. 1** Typical UAV communication scenario [XX].

The main contributions of this article are as follows:

- Review the operation of each UAV to identify malicious UAVs

- Use the hash function to detect fake packets
- Use the hash function to interact between normal UAVs in the network
- Use UAV behavior information to prevent fake information from being sent to other UAVs
- Use a mobile agent based on a three-step negotiation process to design a secure mechanism

The rest of the article is organized as follows:

In section 2, Security attacks and detection schemes are described. Section 3 of this article describes the Hash Function Algorithm (HFA) in great detail. In section 4, the proposed SAUAV schema with its phases is fully described. Section 5 of this article presents the Experimental Results and Performance Evaluation. Finally, the Conclusion and Future work are given in the last section.

## 2 Security attacks and detection schemes

We discuss the issues of cyber security risks targeting UAVs and detection outlines providing safety for the UAV in the following section. This paper lists all the symbols and abbreviations used in Table 1.

**Table 1** Acronyms and notations.

| Symbol | Description | Symbol | Description |
|---|---|---|---|
| $UAV$ | Unmanned Aerial Vehicles | RTH | Return-to-home |
| SAUAV | Secure Agent Unmanned Aerial Vehicles | SVM | Support vector machine |
| $NS-3$ | Network Simulator 3 | STL | Self-Taught Learning |
| $IDS$ | Intrusion Detection System | UAVJ | UAV jammer |
| $GH$ | Gray hole | IRs | Information receivers |
| $BH$ | Black hole | SNR | Signal-to-noise ratio |
| $SH$ | Sinkhole | GTs | Ground Terminals |
| $FP$ | False positive | SCA | Successive convex approximation |
| $FN$ | False negative | IB | Identity Based |
| $TP$ | True positive | SHA-3 | Secure Hash Algorithm 3 |
| $TN$ | True negative | ENCR | Encryption |
| $DR$ | Detection rate | AUT | Authentication |
| PDR | Packet Delivery Rate | SG | Signature Generation |
| RE | Residual Energy | SV | Signature Verification |
| SFA | Security Framework Aircraft | $Ti_A$ | Trusted Authority |
| HFA | Hash Function Algorithm | $Pi_{RTA}$ | Private key |
| RREQ | Route Request | $P$ | Random associated number |
| RREP | Route Reply | $K$ | Identity of source |
| Sinkhole | SH | $T$ | Time stamp |
| Blackhole | BH | Grayhole | GH |

### 2.1 Security Attacks

Because UAVs exchange information over a wireless channel, they are exposed to threats and cyberattacks, including network layer attacks such as sinkhole by intruders. The attacks studied in this study to prevent their subversive activities are listed in Figure 2.

- *Sinkhole Attack*: In this type of attack, the intruder UAV tries to attract network traffic by advertising fake routes. Used this type of attack to perform selective send attack and change routing information [5-7].
- *Black Hole Attack:* It is a type of denial of service attack that receives malicious UAV packets but does not deliver them to the destination UAV or neighboring UAV and removes them [8-10].
- *Grey Hole Attack:* It is a type of denial of service attack that receives malicious UAV packets and selectively delivers a percentage of the packets to the destination UAV or neighboring UAV and deletes the rest of the packets [11-13].

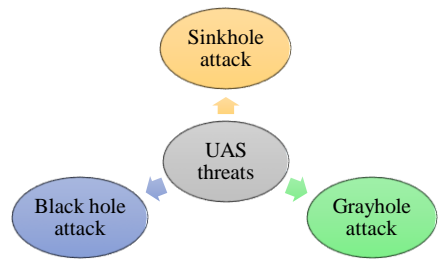

**Fig. 2** UAV cyber security threats

## 2.2 Detection schemes

Nowadays, establishing security against malicious nodes in routing protocols of UAV has drawn a lot of attention. In the following, we present some of the studies conducted on cyber security attack detection.

An authentication-based security approach called HetNets is introduced in [12], inspired by Identity (IB) to secure military UAVs. In this way, confidential information can be overheard from the enemy environment. This method can also be used for civilian applications. The simulation results showed that the proposed method can be resistant to forgery, replay, message change.

In [13], a reliability-based method is proposed to cover the performance of UAV networks with orthogonal and non-orthogonal multiple access capability. In this method, directional modulation tool is used to improve the security performance of the UAV network. Stochastic geometry has also been used to extract the parameters of effective stealth power, stealth disconnection probability and connection disconnection probability. Then, to improve the stealth performance, a genetic algorithm was used to improve the active antenna. Limited energy users with low rate requirements and the probability of covering energy information have been calculated using OMA and NOMA methods.

A secure communication system is proposed in [14] in which the sender UAV sends packets to the destination UAV quite intelligently. UAV uses radio waves to prevent fraud, eavesdropping. Game theory has been used to defend against wiretapping. The Q learning algorithm is also used to achieve an adaptive policy.

A method based on alternating optimization and sequential convex approximation to maintain the security of UAVs in communication between UAVs is proposed in [15]. In this method, the source UAV, while sending information to GN, creates noise so that the malicious UAV cannot hear it. In this method, the path between UAVs is optimized using dense / dense power allocation.

This method uses a set of specific rules to detect malicious UAVs. This method prevents attacks of spreading false information, falsifying information, blocking and grayhole and blackhole attacks. The simulation showed that the efficiency of this method is potentially superior to the false positive and false negative criteria. It also blocks intrusive UAVs without any communication overhead [16].

In [17], a cybersecurity threat model, called CST-UAS, is presented based on the various security attacks that have been analyzed for a UAV system. This model shows the possible routes of the attack. Due to the confidentiality of the information needed to identify threats, it is still difficult to determine which threats may affect UAV systems the most. This article identifies high priority threats and applies and eliminates those related techniques. The model will also help researchers researching UAV systems understand the threat characteristics of UAVs to allow them to address various system vulnerabilities. The evaluation results of this article are shown in tables in the article.

In [18], a security framework is proposed to protect UAVs against malicious behavior that targets UAV communication systems. In addition, the challenges of cyber detection methods for the security of UAVs have been investigated. Also, in this article, we provide a detailed review of the various cyber identification schemes in UAVs and highlight several cyber-attacks that can occur on this network. Numerical results show that the CS-AVN method has a high prediction and accuracy compared to the related methods previously performed.

In [19], a clustering method called HVCR is proposed to accelerate the convergence rate of the formation of several UAVs. This article has been reviewed and analyzed from two aspects of cooperation control and secure communication. In terms of secure communications, the virtual communication loop strategy has used hierarchies to reduce the boundaries of group communications and the least insecure domain. From the aspect of joint cooperation control, by adjusting the flight control coefficient to accelerate the convergence of several UAVs, the UAV group forms a herd. The output of the results of this paper showed that the efficiency of the HVCR algorithm is superior to the communication criteria.

In [20], the concept of ground station is used. In this paper, by jointly optimizing the route and transmission power of the UAVs, as well as user planning, they maximized the minimum amount of secrecy among ground users. This method examines multiple UAVs simultaneously in which one UAV wants to send confidential messages to multiple terrestrial users. In this system, to effectively solve this non-convex problem, we use sequential high-level minimization techniques that address a sequence of approximate convex problems for each block of variables. To improve security performance, a cooperative UAV is also provided that transmits the blocking signal. The simulation results showed that the BSUM-based method works better than the basic methods.

## 3  Hash Function Algorithm (HFA)

Exhibiting the equivalence in the equation $H(m1) = H(m2)$, where m1 and m2 are two distinct input messages is troublesome through employing hash functions, since in a hash function, the delay of the processing speed must be minimised so that it can be computationally efficient [18]. A next generation of standard in security employed in electronic communications is SHA-3, which transforms the digital messages into "message digests" to register digital signatures. To facilitate the detection of modifications in the message originally sent, changing the original message modifies the message digest. The detailed necessities of the algorithms for secure routing as well as the crucial management services to achieve encryption(encr), authentication(aut) and mechanism for digital signature(DS) are presented in the subsection. This algorithm generates digital signature for any information given, transmitted between the source UAVs and the destination UAVs. In general form, the ECDSA is implemented in three phases:
1) Key pair generation,
2) Signature Generation (SG), and
3) Signature Verification (SV).

It should be noted that phases 1 and 2 are performed in the source UAV, while the destination UAV carries out the last phase.

### 3.1 Registering with Trusted Authority

As demonstrated in Eq. (1), all the UAVs present in the network must select a random point, e.g. $Pi_R \in Zi_P^*$, to generate a private key and calculate the public key, e.g. $Pi_{UTA}$, via multiplying the private key with point generator, to perform registration.

$$Pi_{UTA} = Pi_{RTA} * Zi \tag{1}$$

For instance, registering to the network for an arbitrary source UAV (A) requires transmitting the identity ($IDi_A$) to the Trusted Authority ($Ti_A$). Once the identity is received, the trusted authority will then calculate the source node identity via multiplying the identity with the trusted source private key ($Pi_{RTA}$). Finally, the trusted authority will transmit back the generated $IDi_A^{'}$, as shown in the Eq. (2), to the source node.

$$IDi_A^{'} = IDi_A * PRi_{TA} \tag{2}$$

Moreover, the identity is checked in the source UAV (A) via multiplying $ID_A^{'}$ with the point generator. Furthermore, the identity of the source UAV is checked in the source UAV as well via multiplying $ID_A$ with the public key from the trusted authority. The positive acknowledgment according to Eq. (3) is replied only if both points are equal.

$$IDi_A^{'} * Zi = IDi_A * PUe_{TA} \tag{3}$$

All the intermediate nodes as well as the destination UAVs must register to the network according to the similar procedure discussed in [18].

### 3.2 Employing EC Cryptography to Authenticate Nodes

Following the registration with trusted authority, a mechanism to authenticate all the registered UAVs is required to secure a network. To perform such authentication, public and private keys must be generated by all registered UAVs (e.g. to authenticate one hop source UAV and destination UAV) according to the ECC algorithm discussed in Sect. 3.2. First, the identity of the trusted authority ($P$) and the identity of source ($K$) are calculated, and are summed up to result in the secure identity information $D^{'}$. Next, the security associated information $A_{sum}(D^{'}, M, U)$ are sent to the destination UAV $D$ from the source UAV. In this security associated information, ($D^{'}$) is the secure identity information, $P$ is a random associated number, and $T$ is the time stamp, where:

$$P = ID_A^{'} * P_A * PUe_B$$
$$K = ID_A * M * PUe_{TA}$$
$$D^{'} = P + K \qquad (4)$$

Once the information $A_{sum}(D^{'}, M, U)$ is received, the target UAV verifies the timestamp. If the timestamp is verified, the validity of the security identity information $B^{'}$ is further checked by calculating $D^{'}$. Moreover, the identity of $A$ ($ID_A^{'}$) is also calculated in the destination node via multiplying the random number with the generator point $F$. If the calculation results for $D^{'}$ and $F$ are equal, the node $A$ will be authenticated, and will be rejected otherwise. In Eq. (5), the complete procedure is demonstrated.

$$D = ID_A^{'} * P_B * PUe_A$$
$$D^{'} = D^{'} - D$$
$$F = ID_A^{'} * M * G \qquad (5)$$

To obtain an authenticated network, all the nodes registered in the network should follow the same procedure in a similar manner for mutual authentication [18].

## 4 The proposed SAUAV schema

A cyber-security threats-immune schema is designed in the following section utilizing the HFA algorithm. The suggested technique contains two phases including an overview of the SAUAV model discussed in Phase 1 and details of SAUAV schema explained in Phase 2. Within the proposed SAUAV, a hybrid solution is presented to protect the unmanned aerial systems that are effective in two aspects: First, it contains high detecting accurateness and low false-negative and positive rates and second, it quickly discovers and isolates attacks. Within the suggested scheme, the security issues are prevented including BH, SH, and GH attacks able to target the UAV. To discover the cyber-attacks with high precision, it is possible to add other properties rather than Table 2.

**Table 2** Cyber-security attacks features

| Cyber security threats | Features |
| --- | --- |
| Sinkhole attack | Data injection rate |
| Blackhole attack | Data injection rate |
| Gray hole attack | Data injection rate |

### 4.1 Phase 1: Overview of the SAUAV model for detecting malicious UAV

The proposed algorithm is implemented on AODV protocol. According to the behavior of the nodes in the UAV network, we try to be able to identify and to eliminate malicious nodes in order to prevent of presenting wrong information to the checker UAV nodes in this algorithm. When the number of malicious nodes is increasing, as a result the number of sending request for comment would be increased. The overhead is increasing by an increase in the number of malicious nodes, because more nodes can start the process of sending request for comment or judgement. The more the overhead, the more the delay is. Therefore, by identifying malicious nodes the overhead and then the delay can be decreased. As the number of malicious nodes increases, the overhead of the algorithm becomes larger, as a result the identification of the malicious nodes becomes more difficult, for this purpose the delivery rate of the data decreases when the malicious nodes increase.

In order to detect the malicious UAV, we utilize the following rules in the proposed SAUAV:

- The UAV node has sent a number of data packet towards other UAV nodes, cannot be the malicious one.
- The UAV node has received many of the data but did not send them back, it may be the malicious one.
- The malicious UAV is one that has sent at least one RREP packet.
- The UAV node has received many of the data but did not send them back, and has sent at least one RREP packet, it is surely the one malicious UAV node.
- The node sending at least one response message of the route to the sender node of the route request earlier than other nodes, may be the malicious UAV node.
- The node that comprises the greatest sequence number and the lowest hop-count in its route response message, may be the malicious UAV node.

The principles of the proposed SAUAV are expressed as follows:

- The data corresponding to the activities of the nodes (number of sending data, number of receiving data, and number of receiving responses) is saved and investigated.
- Regarding the node of a neighbor that has sent at least one RREP packet, the requesting packet for comments among neighbors could be sent.
- The data saved in neighboring nodes regarding the sender node of RREP packet are received.
- The received data and declaration for the comment associating with the UAV node which is being malicious, are investigated.
- A packet of warning message was sent to quarantine the malicious UAV node and it`ll be distributed across the network.
- The UAV nodes under quarantine would be eliminated during routing process.

**4.2 Phase 1: Details of the proposed SAUAV for purging malicious UAV from network**
In this section, the Secure AODV focused on SAUAV which is based on smart agent's use. We used smart agents to aware the nodes from their valid neighbors to avoid of listening to the data generated by the malicious nodes. We will also explain the description and definitions regarding designing of agents and structure of memory in nodes, and how to use smart agents.

*4.2.1  Designing of an agent*
The smart agents are new and smart samples for distributed applications, and they are able to do their duties instead of human.  In fact, a smart agent is a programmatic or executor code that migrates among nodes as a form of agent packets. The format of the agent packets will be discussed in the next section. Following the paper, we will use agents and changeable packets of agents. There are two tangible and major differences among agents and other existing solutions that we will express them in the proposed method. First of all, in contrast to some of the previous methods, we only use one agent to identify the malicious UAV nodes. Second, the agents are not associated with each other, and instead they are in fact associated with each other through the fixed nodes.

Therefore, the agent programs will be accelerated, as a result computational costs also are going to be decreased, and extra overheads won`t be existed in the agents` communications. Such optimization results in decrease in the energy consumption among nodes.

**The format of the agent packets:** The agent packet is encapsulated in an agent packet object as shown in (Table 3). As we use only one type of agent, therefore there is no need to save extra fields for type of an agent and their communicational style. Both the fields of the source node and the target node number are utilized for storing the source and target node identifier in a move, respectively. The agent program comprises a hash function -which is called "hash function of an agent"-, and two unique codes - which is called "code 3 and data code". As code 3 is an output of different hash functions executions (e.g. Hash function of node), a copy of them (hash function of node and code 2) is stored in the node memory as a different unique code -which is called "code 2". A hash function of agents and unique codes (e.g. Code 3) are utilized for interacting with valid nodes in the network and to create an availability set for sections corresponding with the data of the agents` packets. The data section consists of two fields: A valid bit and an agent bit. As these and those fields storing in the nodes are the same, we will explain them in the section of the node memory structure in more detailed. If the data section existing in the text is simple, an aggressor can easily use the data. Therefore, data code is used to encode the data and to ensure integrity of the data section in an agent packet. If an identified agent is a reliable node, then it sends the data code towards the node. Hence, the node can extract the correct data from agent packets by reducing the volume of the data section from data code. Nevertheless, if a reliable node seeks for adding new data to the data section, first of all it has to add them to the data code, and then, to interpolate them inside the agent packets. The data code and the data section have both the same size (e.g. 1 bit). The upper and lower boundary of the data section are used for storing a valid bit and an agent bit, respectively. At the time being, as an aggressor is not able to limit the smart agents, so it can not to access to the data code and real data within the data section. Moreover, an aggressor is not permission to fill a part of data with fake values. Finally, an object of agent packet provides some of the auxiliary methods to communicate with the nodes. The agent packet format used in this paper are provided Table 3.

**Table 3** Agent packet format.

| Agent packet | No of source UAV | No of destination UAV | Code 3 | Hash function output | Data code | data |
|---|---|---|---|---|---|---|

**Agent migration:** In our solutions, migration means to move an agent from one agent node to the one-hop neighbor's node and to come back to its main node. Regarding this topic, there is no need to store a route for an agent migration within agent packet, since a smart agent moves only towards the one-hop nodes and comes back to its main own node. An agent node is a valid and general node which maintains a smart agent. In other words, an agent node accounts for a general node receiving an agent packet from its neighbor. The other definitions provided by us is term cycle of an agent referring to it`s all one-hop neighbors as an agent migration as well as an agent node. In our solution, after a smart agent is placed on a node, then term cycle of an agent is randomly performed per 5-10 seconds. In addition, if two or Multi-agents deal with each other at the same time, the node will maintain one of them which had recently received with a little difference in time, and will ignored the rest. For this purpose, if an agent within a migration did not come back to its initial node after a particular time, the neighbor node won`t be immediately account for an aggressor;

instead the given agent will be twice re-sent in the randomized periods. If an agent did not come back to its initial node afterwards, then it will be known as an aggressor.

*4.2.2 The node memory structure*

In contrast with other methods, we only maintain one-hop neighbors of a node in a table called "neighboring matrix" to store memory consumption within the node. The data stored in neighboring matrix included: the number of the node, a valid bit, and an agent bit. The number of the node is the number of one-hop neighbor node identification. A valid bit is utilized for determining reliable neighbors, the valid nodes within Secure AODV only receive or send the data from reliable nodes, and never communicate with the malicious nodes. As the smart agents are responsible for determining the reliability of a node, the agent nodes, in order to receive secure data only from the smart agents, have to be recognized. As a result, an agent bit is used for identifying the agent nodes. Before spreading the nodes in the environment, there are two unique codes in this regard: The code 1 and 2 are entering into their memory. The code 1 is not encoded, yet but it is in a simple case. While, code 2 within code 1 accounts for an output of an agent hash function execution (e.g. an agent hash function). The nodes were also equipped using hash function called "the node hash function" in which it is varying from those stored in the agents` program. These two codes and hash function are usable for detecting fake agents, and interaction with the valid nodes within network. Ultimately, the smaller section of the memory is used for habitation of roaming smart agents.

*4.2.3 The Algorithm*

Our proposed SAUAV is consisted of two steps for identifying and preventing of attacks in the following: The network deployment step illustrates how the network is being configured, while the UAS network maintenance step is an indicator of how the network security is maintained.

**Deployment phase of the network:** At the beginning, the nodes are uniformly distributed in the UAV network. Consequently, the neighbor randomly selects a number of the UAV nodes from smart agents according to the desired percentage to send the agent packets. When a node receives an agent packet from its neighbor that is known as an agent node. Following this step, all nodes spread the HELLO packets for finding existing nodes in the radio frequency range as well as creating neighboring matrix. It is completely clear that the malicious nodes can convert themselves into neighboring matrix. After finding neighbors process, each input within neighboring matrix encompasses identifiers of one-hop neighbor nodes, but an agent bit and the valid bit remain incorrect, yet. There is a wireless network as shown in the Figure. The Figure 3, in which the nodes A & H show an agent, and E accounts for a malicious node, while the Table 4 indicates the nodes A & H, Then the node B after sending HELLO packets becomes neighboring matrix. As shown in the Figure, during the present step, all neighboring nodes are found in which range consists of malicious cases.

**Maintenance phase of the network:** After deployment phase of the network, the agents commence with an agent cycle. But, before presenting any information to a node or receiving from that, a three-step negotiation known as (the process of confidence-building) is performed among an agent and the node. If a node is reliable, the interactions would be commenced; otherwise, the given node accounts for an aggressor. If, the neighbor node is reliable, then valid bit related is correctly altered in neighboring matrix (for instance; one) when an agent returns to the main node (or an agent node), and Otherwise, it remains incorrect (e.g. zero); Moreover, if the neighbor node is an agent node, as a result it`s agent bit remains true (true). As only a percent of the nodes comprises an agent (like; agent nodes), the agent nodes send a packet of (confidence packet) to it`s all reliable neighbors after determining the reliability among neighbors, and tries to aware the agents of reliable and malicious nodes. Additionally, as the nodes are moving, the received signal strength is calculated when an agent returns towards an agent node. If it is lower than one threshold, the neighbor node is then eliminated from neighbors' list, but it is assumed that the nodes are moving. If the eliminated agent bit of the node is incorrect, an agent node sends a control packet for re-performing of seeking for neighbor process, because it has no agent and by no means can be covered by smart agent.

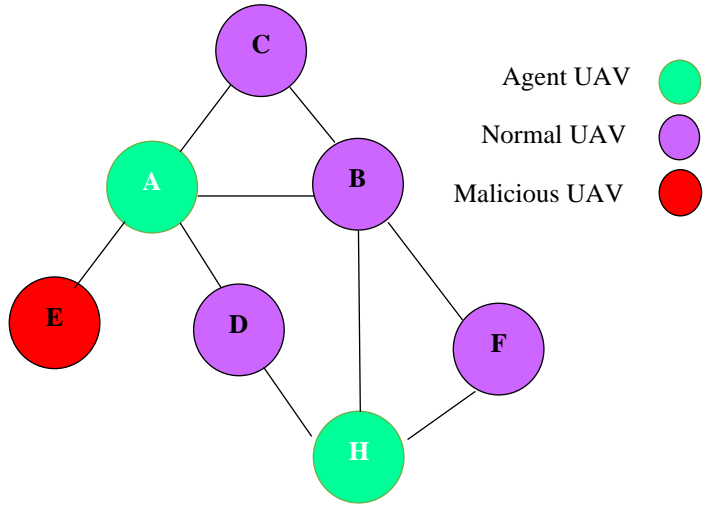

**Fig. 3** UAV with agent UAV, normal UAV, and malicious UAV.

**Table 4** the $UAV_A$, $UAV_B$, $UAV_H$ are the neighboring matrix after sending HELLO packets.

| | $UAV_{ID}$ | $UAV_B$ | $UAV_C$ | $UAV_D$ | $UAV_E$ |
|---|---|---|---|---|---|
| $UAV_A$ | Valid | 0 | 0 | 0 | 0 |
| | Agent | 0 | 0 | 0 | 0 |

| | $UAV_{ID}$ | $UAV_A$ | $UAV_C$ | $UAV_F$ | $UAV_H$ |
|---|---|---|---|---|---|
| $UAV_B$ | Valid | 0 | 0 | 0 | 0 |
| | Agent | 0 | 0 | 0 | 0 |

| | $UAV_{ID}$ | $UAV_B$ | $UAV_F$ | $UAV_D$ |
|---|---|---|---|---|
| $UAV_H$ | Valid | 0 | 0 | 0 |
| | Agent | 0 | 0 | 0 |

**Table 5** The $UAV_A$, $UAV_B$, and $UAV_H$ are the neighboring matrix after an agent migration.

| | $UAV_{ID}$ | $UAV_B$ | $UAV_C$ | $UAV_D$ | $UAV_E$ |
|---|---|---|---|---|---|
| $UAV_A$ | Valid | 1 | 1 | 1 | 0 |
| | Agent | 0 | 0 | 0 | 0 |

| | $UAV_{ID}$ | $UAV_A$ | $UAV_C$ | $UAV_F$ | $UAV_H$ |
|---|---|---|---|---|---|
| $UAV_B$ | Valid | 1 | 0 | 0 | 1 |
| | Agent | 1 | 0 | 0 | 1 |

| | $UAV_{ID}$ | $UAV_B$ | $UAV_F$ | $UAV_D$ |
|---|---|---|---|---|
| $UAV_H$ | Valid | 1 | 1 | 1 |
| | Agent | 0 | 0 | 0 |

Table 6 illustrates the neighbors' matrix for the $UAV_A$, $UAV_B$, $UAV_H$ just after an agent migration.

**Table 6** The $UAV_B$ neighboring matrix after receiving reliable packets from the $UAV_A$, $UAV_H$.

| | $UAV_{ID}$ | $UAV_B$ | $UAV_C$ | $UAV_D$ | $UAV_E$ |
|---|---|---|---|---|---|
| $UAV_B$ | Valid | 1 | 1 | 1 | 1 |
| | Agent | 1 | 0 | 0 | 1 |

The bits of an agent and the valid are filling with the relating values. But, as the node B has no agent, it will not enjoy any information about the $UAV_C$, $UAV_F$. So, the agent nodes (here $UAV_A$, $UAV_H$) send the confidence packets towards their reliable neighbors (as B). After receiving confidence packets, the $UAV_B$ updates it`s matrix, as shown in (Figure 4). Therefore, all nodes are aware of all of their one-hop neighbors, and the malicious nodes are being well identified in this regard. According to this case, the nodes receive/send routing data and data packets from their reliable neighbors as well, and they will not listen to the aggressors. A significant point including (the network longevity, energy level of the nodes, special agent nodes), if it goes away resulting in their demolition, as a result we designed one scheme for this task very well. When the energy between the general nodes is lower than the threshold, the neighbors will deal with a death packet, and then the death node is eliminated of the neighbors' list. But there is a little difference for the agent nodes. When an agent node is dying, it sends it`s agent to one of the reliable neighbors having no agent, and then acts as the general nodes. So, an agent is placed on the new node and operates like this.

*4.2.4 Detection malicious attacks*

We explain how a SH UAV is being identified using codes (code 1, 2, and 3). Regarding movement of the nodes, the maintenance stage of the network has to be periodically performed during network longevity. Therefore, a malicious node is able to fake the identification of a valid node, it also can be placed in the list of valid neighbors of the nodes in which tries to configure neighboring matrix. As an aggressor does not know when a valid node wants to perform reconfiguration (means to update neighboring matrix), hence it has to permanently listen to the network traffic, so for this purpose this can lead to the rapid decrease in its energy, and it is immediately getting lost. As recently mentioned, before that a node attempting to send some data to a mobile agent, the node and the agent need to be confidence of each other. This is a confidence method as shown in Figure 5. As shown in Figure 5, a valid node includes main codes like; code 1, and code 2 while a valid agent has only code 3. In fact, we use this type of method in order to identify enemy nodes. The confidence method is expressed as follows:

As shown in Figure 4, after that an agent is going to be placed in the node, the agent then sends a request for code 1, and code 2 according to its unique hash function (e.g. Agent hash function) which is considered as stage 1, afterwards the agent sends code 2 to the node at the next stage. If it is matched with the code (2) stored in the node, it may trust to this agent, and also may conduct (for instance; node hash function) the code 3 using it`s unique hash function regarding code 2. It is the time for an agent to trust the node. In the last stage of the process of confidence, the code 3 is sent towards an agent. If it is equaled to a code stored in the source code of an agent, the agent then trusts to the node and sends the data code towards it. At the time being, there are three situations regarding confidence method:

**A valid agent inside a valid node:** In this situation, after that an agent was placed in the node, it is able to send a request for code 1, and code 2 is conducted as more recently expressed. Consequently, the representative sends the code 2 towards the node at the next stage. As the code 2, and the other one sent and stored inside the node are the same, the node will trust an agent. At the next stage, the valid node generates code 3 based on code 2, and then can send it towards an agent. Because of this code equals to the other one stored in an agent program, an agent would trust this node, and will send its data code towards it. Then, the node can extract the correct data from an agent packet or is able to add new data to that.

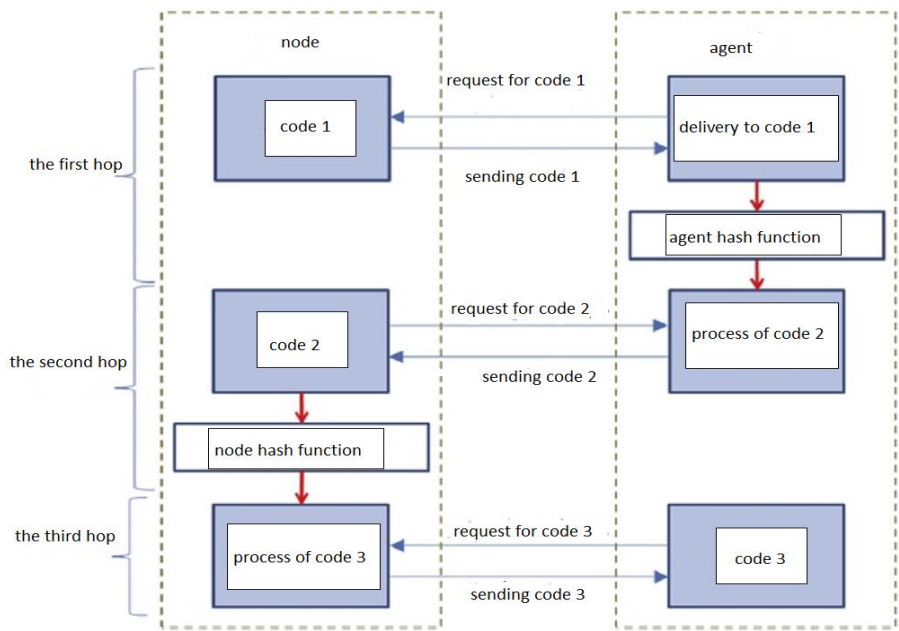

**Fig. 4** The process of confidence from a smart Agent and the UAV.

**A valid agent inside the enemy UAV:** Similar to the previous situation, the agent requests for code 1, and then generates the code 2 using the agent hash function as well as sending it towards the UAV, and ultimately waits for delivering code 3. But the UAV is unable to send such correct code, because of the lack for the main code 1, and the method of UAV hash function. Thus, after that the agent received incorrect code 3, it will not trust again the given UAV and do not send its data code any way. As a result, an aggressor UAV can`t mine or change the correct data from agent packets` data section.

**A fake agent inside a valid UAV:** In such circumstance, after that the fake agent was placed in such UAV, and when received code 1, then it won`t be able to present the correct code 2 for the UAV at the next stage, because this type of agent has no main method of the agent hash function. As such code is not the same was stored in the UAV, therefore the UAV does not trust the agent, and the agent is consequently ignored. Additionally, an aggressor UAV may involve the agent with some difficulty instead of returning it to the initial UAV or to conduct it towards wrong routes. In this regard, if an agent did not return after a particular time (according to migration`s definition), the neighbor UAV is then considered as an aggressor. The flowchart of proposed SAUAV is given in Figure 5.

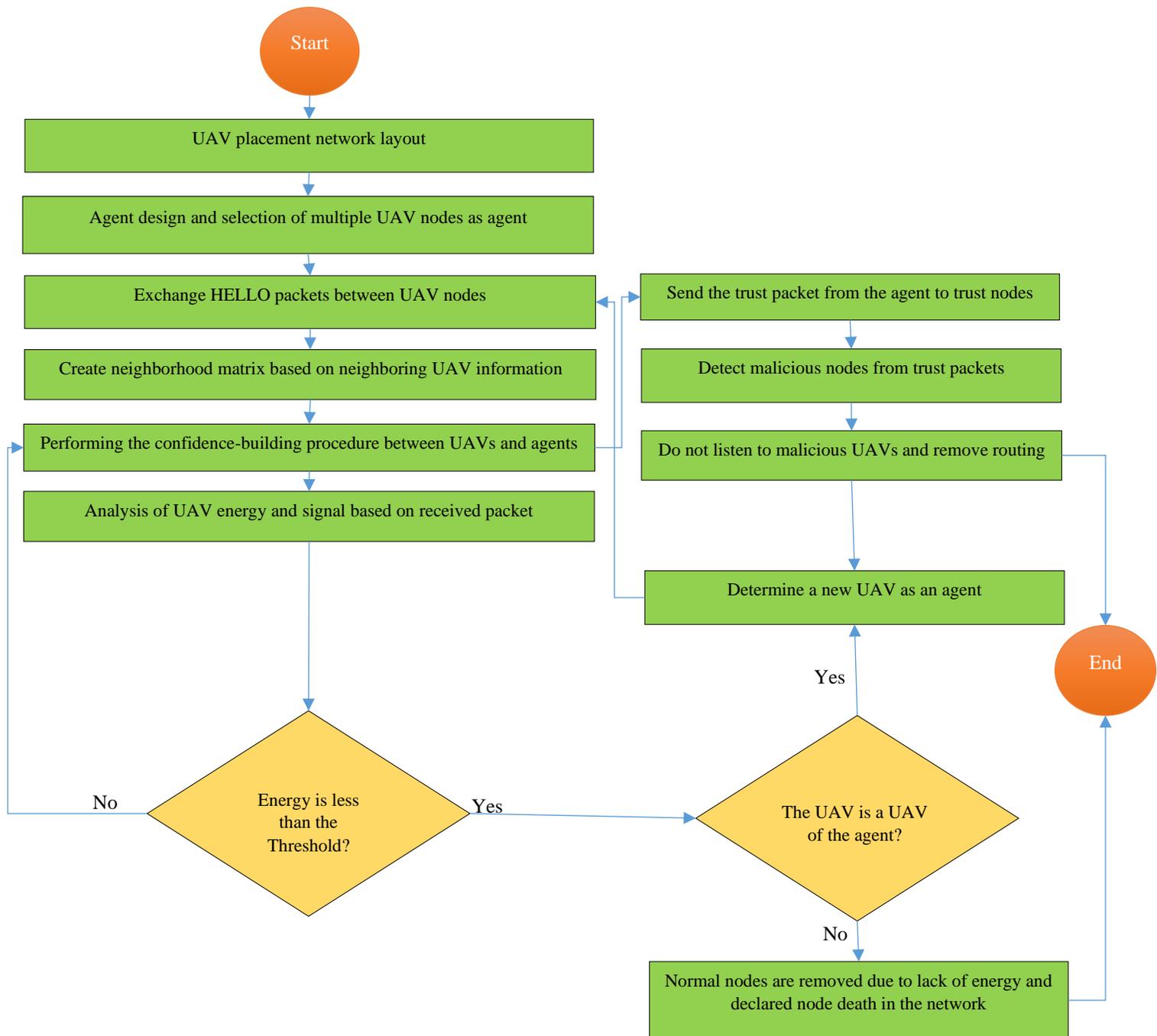

**Fig. 5** Flowchart of the SAUAV.

## 5   Experimental Results and Performance Evaluation

In this section, experimental results and performance evaluation, the proposed method and 4 other methods are discussed.

## 5.1 Performance metrics

This section analyses security parameters and evaluates the qualitative performance in the form of numerical results to validate the performance of the proposed SAUAV method. To demonstrate a feasibility study, the performance analysis of SAUAV has been divided into four parts:

1) PDR,
2) FP,
3) FN,
4) DR,
5) RE

Therefore, to show the superiority of the proposed method, it was compared with four methods SAUAV, BSUM-base, HVCR, CS-AVN and CST-UAS under the criteria of DR, FN, FP, PDR, and RE.

**Table 7** Variables used in PDR criteria according to Eq. (6)

| Variable | Description |
| --- | --- |
| $X_i$ | Total packets received by UAV with $i$ identification number. |
| $Y_i$ | Total packets sent by UAV with $i$ identification number. |
| $n$ | Iteration |

**PDR:** The number of packets successfully received by the destination UAV is divided by the number of packets sent by the source UAV [21-24]. This criterion is shown in Eq. (6).

$$PDR = \frac{1}{n} * \frac{\sum_{i=1}^{n} X_i}{\sum_{i=1}^{n} Y_i} * 100\% \tag{6}$$

**FP:** Number of UAVs mistakenly identified as malicious UAVs. This criterion is shown in Eq. (7).

$$FP = \left(\frac{FP}{FP+TN}\right) * 100 \tag{7}$$

**FN:** The number of malicious UAVs that were mistakenly identified as normal UAVs is divided by the total number of malicious UAVs in the UAV network [25-27]. This criterion is shown in Eq. (8).

$$FN = \left(\frac{FN}{FN+TP}\right) * 100 = 1 - TP \tag{8}$$

**DR:** The number of malicious UAVs detected by smart agents in the proposed method is divided by the total number of malicious UAVs in the UAV network [28, 29]. The *DR* is shown in Eq. (9).

$$DR = \left(\frac{TP}{TP+FN}\right) * 100 \tag{9}$$

## 5.2 The simulation environment

The proposed method was simulated on the Ubuntu 14.04 LTS platform. The NS-3 tool was used as a simulation tool. The simulation environment was considered with dimensions of 2000 by 2000 meters.

## 5.3 Simulation results

In this section, the simulation results are shown according to Tables 10 to 14 and diagrams 6 to 10. Table 8 summarizes the parameters used in the simulation.

**Table 8** Parameters used.

| Parameters | Value |
|---|---|
| Operating system | Ubuntu 14.04 LTS |
| Simulator tools | NS-3 |
| Channel type | Channel/Wireless channel |
| MAC Layer | MAC/802.11. b |
| Traffic type | CBR |
| UAV speed | 180 m/s |
| Layer of Transmission | UDP |
| Size of packet | 512 Byte |
| Malicious rate | 10%, 20%, 30% |
| Type of intruder UAV | SH, BH, GH |
| Range of Transmission | 30 M |

The parameters used in the three scenarios are given in Table 9.

**Table 9** Parameters used for four scenarios.

| Scenario #1 | | Scenario #2 | |
|---|---|---|---|
| Number of UAVs | 500 | Number of UAVs | 500 |
| Malicious UAV rates | 10% | Malicious UAV rates | 20% |
| Time | 1000 | Time | 1000 |
| **Scenario #3** | | | |
| Number of UAVs | 500 | | |
| Malicious UAV rates | 30% | | |
| Time | 1000 | | |

**Table 10** *DR* (30% malicious UAV of overall UAVs) of various approaches with varying degree of UAVs

| Number of UAVs | DR (%) | | | | |
|---|---|---|---|---|---|
| | CST-UAS | CS-AVN | HVCR | BSUM-based method | SAUAV |
| 50  | 58.32 | 67.1 | 68   | 70 | 81.06 |
| 100 | 58.97 | 68.2 | 69.6 | 71 | 81.34 |
| 150 | 59.25 | 68.8 | 70   | 72 | 82.5  |
| 200 | 60.5  | 69.3 | 71   | 73 | 83    |
| 250 | 61.5  | 69.7 | 72   | 74 | 84.3  |
| 300 | 62.6  | 70.3 | 73   | 75 | 85.5  |
| 350 | 63.4  | 71.5 | 73.5 | 76 | 86.9  |
| 400 | 64.6  | 73.5 | 74.3 | 77 | 87.6  |
| 450 | 65.4  | 74.8 | 75   | 78 | 88.09 |
| 500 | 66.8  | 75.9 | 77   | 79 | 89.76 |

**Table 11** *FN* (30% malicious UAV of overall UAVs) of various approaches with varying degree of UAVs.

| Number of UAVs | FN (%) | | | | |
|---|---|---|---|---|---|
| | CST-UAS | CS-AVN | HVCR | BSUM-based method | SAUAV |
| 50  | 24 | 25 | 22 | 20 | 13   |
| 100 | 26 | 26 | 24 | 21 | 14   |
| 150 | 29 | 27 | 25 | 23 | 15   |
| 200 | 31 | 28 | 26 | 24 | 16   |
| 250 | 34 | 30 | 27 | 25 | 17   |
| 300 | 37 | 31 | 29 | 27 | 17.7 |
| 350 | 38 | 32 | 30 | 28 | 18   |
| 400 | 39 | 35 | 32 | 29 | 19   |
| 450 | 41 | 36 | 34 | 30 | 20   |
| 500 | 43 | 37 | 35 | 32 | 21   |

**Table 12** *FP* (30% malicious UAV of overall UAVs) of various approaches with varying degree of UAVs.

| Number of UAVs | FP (%) | | | | |
|---|---|---|---|---|---|
| | CST-UAS | CS-AVN | HVCR | BSUM-based method | SAUAV |
| 50  | 22 | 18 | 16 | 15 | 14   |
| 100 | 25 | 20 | 19 | 17 | 16   |
| 150 | 26 | 24 | 20 | 19 | 18   |
| 200 | 27 | 26 | 23 | 22 | 19   |
| 250 | 31 | 29 | 25 | 23 | 19.5 |
| 300 | 35 | 31 | 27 | 26 | 20   |
| 350 | 38 | 33 | 29 | 28 | 21   |
| 400 | 41 | 36 | 31 | 30 | 22.1 |
| 450 | 42 | 39 | 32 | 32 | 22.6 |
| 500 | 44 | 41 | 34 | 34 | 23   |

**Table 13** *PDR* (30% malicious UAV of overall UAVs) of various approaches with varying degree of UAVs.

| Number of UAVs | PDR (%) | | | | |
|---|---|---|---|---|---|
| | CST-UAS | CS-AVN | HVCR | BSUM-based method | SAUAV |
| 50 | 63.2 | 66.8 | 69 | 71 | 76.56 |
| 100 | 63.6 | 67.3 | 70 | 72 | 76.9 |
| 150 | 64.2 | 67.38 | 71 | 73 | 77.45 |
| 200 | 64.6 | 68.6 | 72 | 74 | 77.99 |
| 250 | 64.9 | 69.34 | 73 | 75 | 78.12 |
| 300 | 65.5 | 69.9 | 74 | 76 | 78.87 |
| 350 | 65.8 | 70.12 | 75 | 77 | 79.5 |
| 400 | 66.4 | 70.54 | 77 | 78 | 80.43 |
| 450 | 66.89 | 70.89 | 78 | 80 | 81.15 |
| 500 | 67.9 | 71.4 | 79 | 81 | 82.5 |

**Table 14** *RE* (30% malicious UAV of overall UAVs) of various approaches with varying degree of UAVs.

| Number of UAVs | RE (%) | | | | |
|---|---|---|---|---|---|
| | CST-UAS | CS-AVN | HVCR | BSUM-based method | SAUAV |
| 50 | 54 | 56 | 60 | 67 | 71 |
| 100 | 55 | 57 | 62 | 68 | 73 |
| 150 | 56 | 58 | 63 | 69 | 76 |
| 200 | 57 | 60 | 63.7 | 70 | 77 |
| 250 | 58 | 62 | 64 | 70.4 | 78 |
| 300 | 59 | 63 | 65 | 70.8 | 80 |
| 350 | 60 | 63.6 | 65.7 | 71 | 82 |
| 400 | 61 | 64 | 66 | 72 | 83 |
| 450 | 62 | 65 | 67 | 73 | 84 |
| 500 | 63 | 66 | 68 | 74 | 85 |

**DR:** As shown in the diagrams in Figure 6, DR increased in all five methods according to three scenarios (first scenario with 10% malicious UAV, second scenario with 20% malicious UAV, and third scenario with 30% malicious UAV), in particular while the number of infiltrating UAVs is high. This increase is less for CST-UAS and CS-AVN methods than other methods. The reason DR is so bad in both CST-UAS and CS-AVN is that they only used the signature-based method to detect GPS position. The HVCR method also uses a clustering algorithm to detect malicious UAVs that cannot completely prevent malicious UAVs. The BISUM-based method uses the block successive upper bound minimization technique, which cannot be effective against malicious UAVs because it uses the least amount of secrecy with joint transmission power optimization. However, in the proposed method, because a mobile agent based on a three-step negotiation process is used to eliminate malicious UAVs, this method can detect UAVs with blackhole, grayhole and sinkhole attack with a detection rate higher than 90%. This result is obtained when the number of normal UAVs and the rate of malicious UAVs are 500 and 30%, respectively. Another reason SAUAV is superior is that it uses mobile agents to inform its normal neighboring UAVs so that they do not care about the data generated by malicious UAVs. Thus, as can be seen from the graphs, in the case where the number of UAVs and the rate of malicious UAV are 500, 10 and 20%, respectively, it is 94 and 92%, respectively.

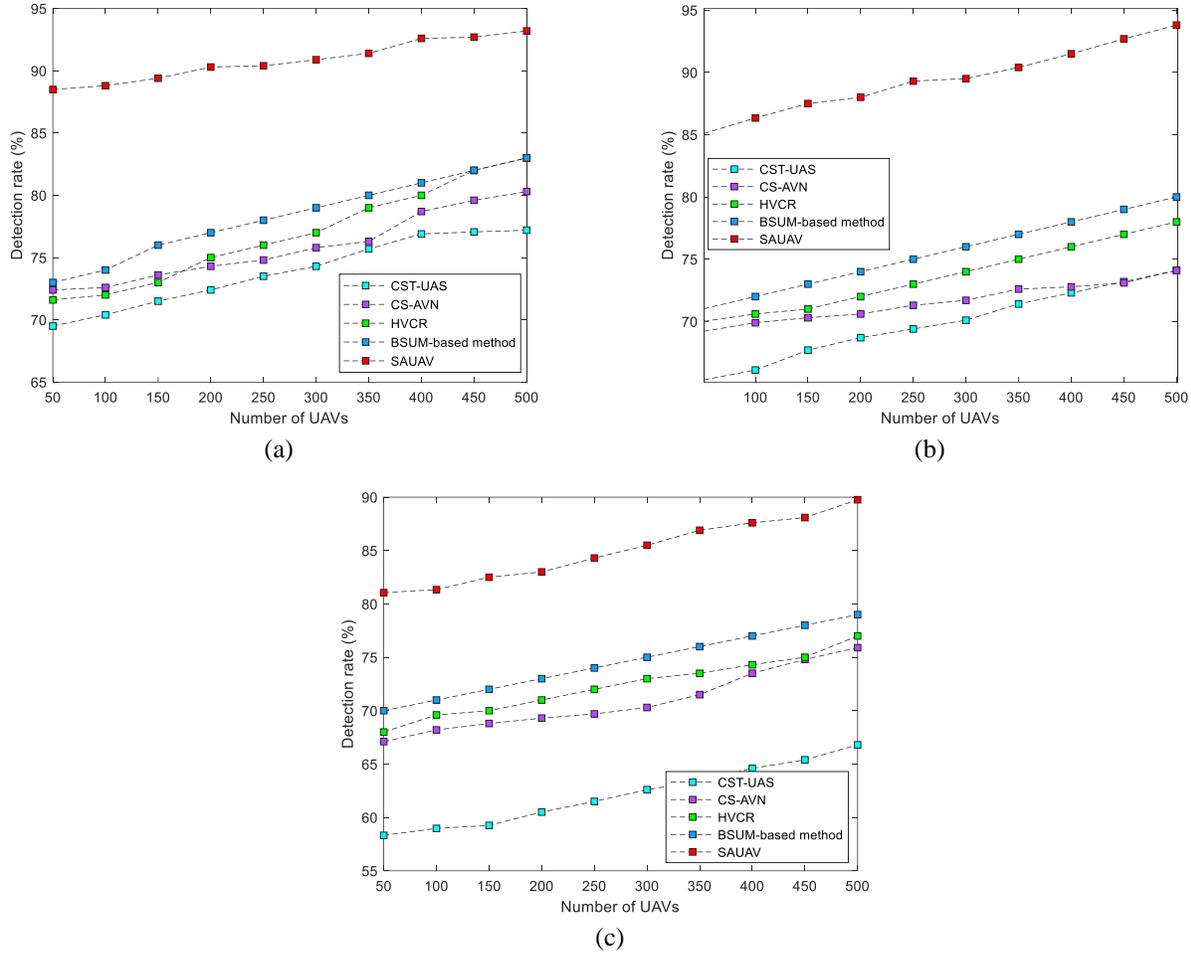

**Fig. 6** Comparison of the SAUAV, BSUM-based method, HVCR, CS-AVN, and CST-UAS models in term of *DR*.

**FN:** As shown in the diagrams in Figure 7, FN was increased in all five methods according to three scenarios (first scenario with 10% destructive UAV, second scenario with 20% destructive UAV, and third scenario with 30% destructive UAV), especially while the number of infiltrating UAVs is high. This increase is greater for CST-UAS methods than for other methods. The reason FN is so bad in both CST-UAS methods is that it only used the signature-based method to detect GPS position. The HVCR method also uses a clustering algorithm to detect malicious UAVs that cannot completely prevent malicious UAVs. The BISUM-based method uses the block successive upper bound minimization technique, which cannot be effective against destructive UAVs because it uses the least amount of secrecy with joint transmission power optimization. But because the proposed method uses two phases to secure communication between UAVs. In the first phase, due to the behavior of malicious UAVs, in order to prevent sending fake information to other normal UAVs, a series of safety rules were used to identify malicious operations to identify malicious UAVs and prevent routing and data transfer operations between other UAVs. Be. In the second phase, a mobile agent based on a three-step negotiation process was used to identify the moving agent of each UAV from reliable neighbors through a three-step negotiation process, to prevent the normal UAV from receiving false information from other malicious UAVs. The combination of these two phases led to secure communications between the UAVs so that packets could exchange information securely. Therefore, according to the diagrams, the output of the proposed method for destructive UAV rates of 10, 20 and 30% is equal to 12, 15 and 20%. But this value is 31%, 35% and 35% for BISUM-based, HVCR and CS-AVN methods, respectively.

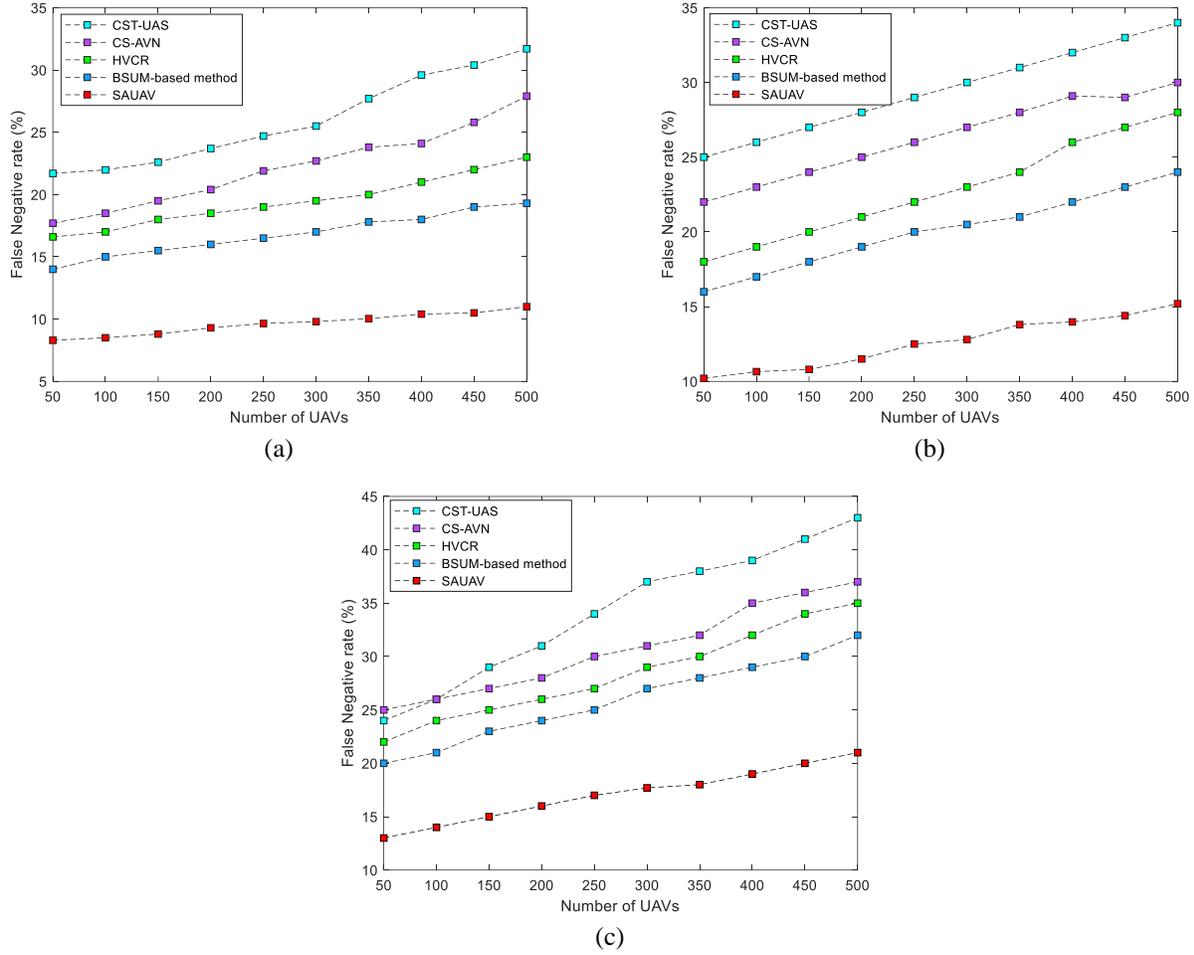

**Fig. 7** Comparison of the SAUAV, BSUM-based method, HVCR, CS-AVN, and CST-UAS models in term of *FN*.

**FP:** As shown in Figure 8, the SAUAV method has a lower *FP* than the other methods. This is in the context that the *FP* of other methods is higher than the proposed method. The reason for the superiority of the proposed method is that, first, it examines the operation of each UAV to identify malicious UAVs. Second, it uses the hash function to detect fake packets. Third, it uses the hash function to interact between normal UAVs in the network, and it also uses a mobile agent based on a three-step negotiation process to design a secure mechanism. The *FP* of all methods is 10, 17, 24, 28 and 33% for SAUAV, BSUM-base, HVCR, CS-AVN and CST-UAS methods, respectively, with a malicious UAV rate of 10%. This output for the second scenario is 22, 25, 30, 35 and 40, respectively. And for the third scenario, it is equal to 24, 34, 34, 41 and 44, respectively.

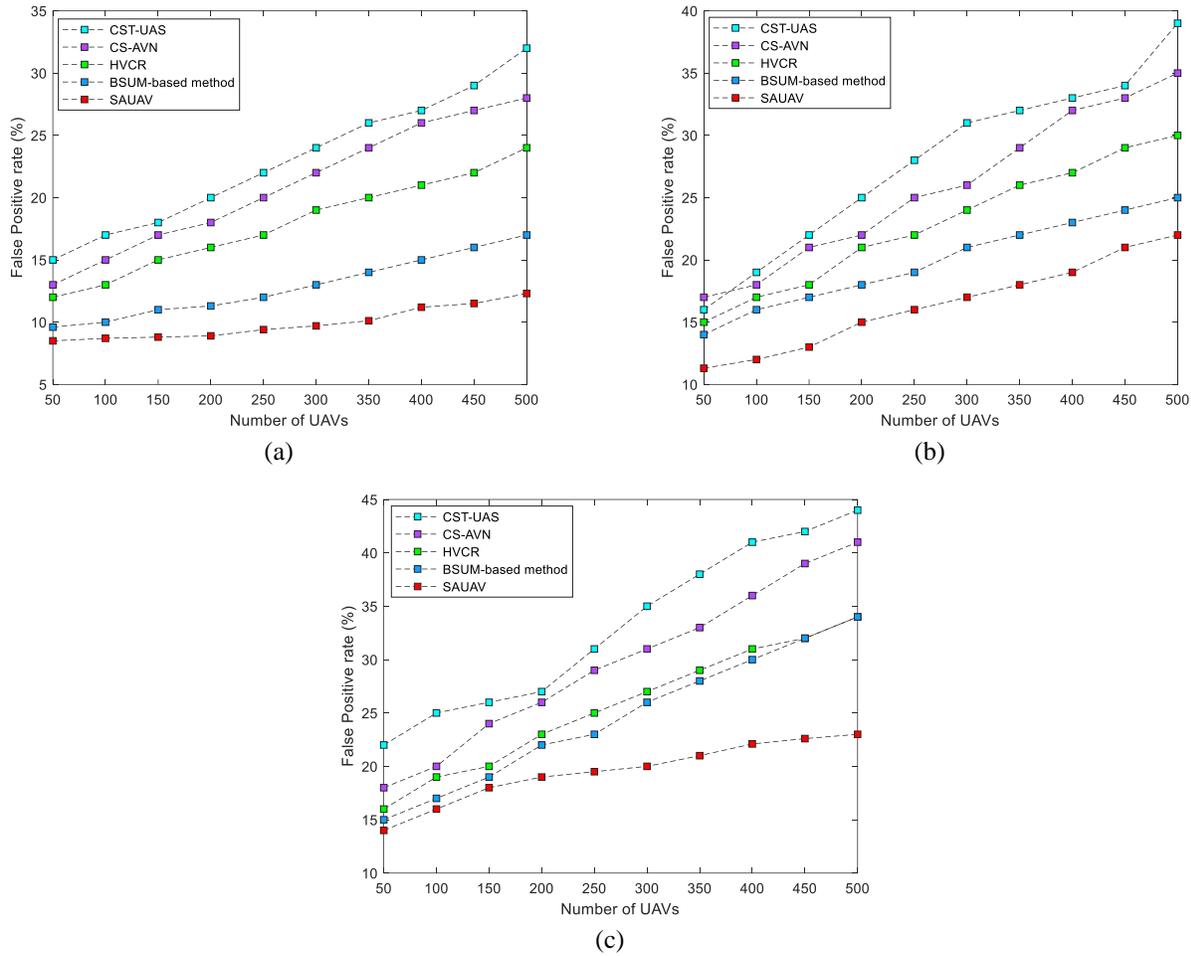

**Fig. 8** Comparison of the SAUAV, BSUM-based method, HVCR, CS-AVN, and CST-UAS models in term of *FP*.

**PDR:** Because malicious UAVs are quickly detected and eliminated from the packet routing and exchange operation cycle, the proposed method performs better than other methods. Also, considering that the proposed method had a higher *DR* than other methods, so it is natural that it has a high *PDR* because malicious UAVs are excluded from routing operations and packet exchange is done by normal UAVs. Therefore, while the destructive UAV rate is 10%, the PDR of SAUAV, BSUM-base, HVCR, CS-AVN and CST-UAS methods are 97, 91, 86, 77 and 72%, respectively. This output for the second scenario is 89, 84, 79, 72 and 71, respectively. And for the third scenario, it is equal to 82, 79, 76, 71 and 67, respectively.

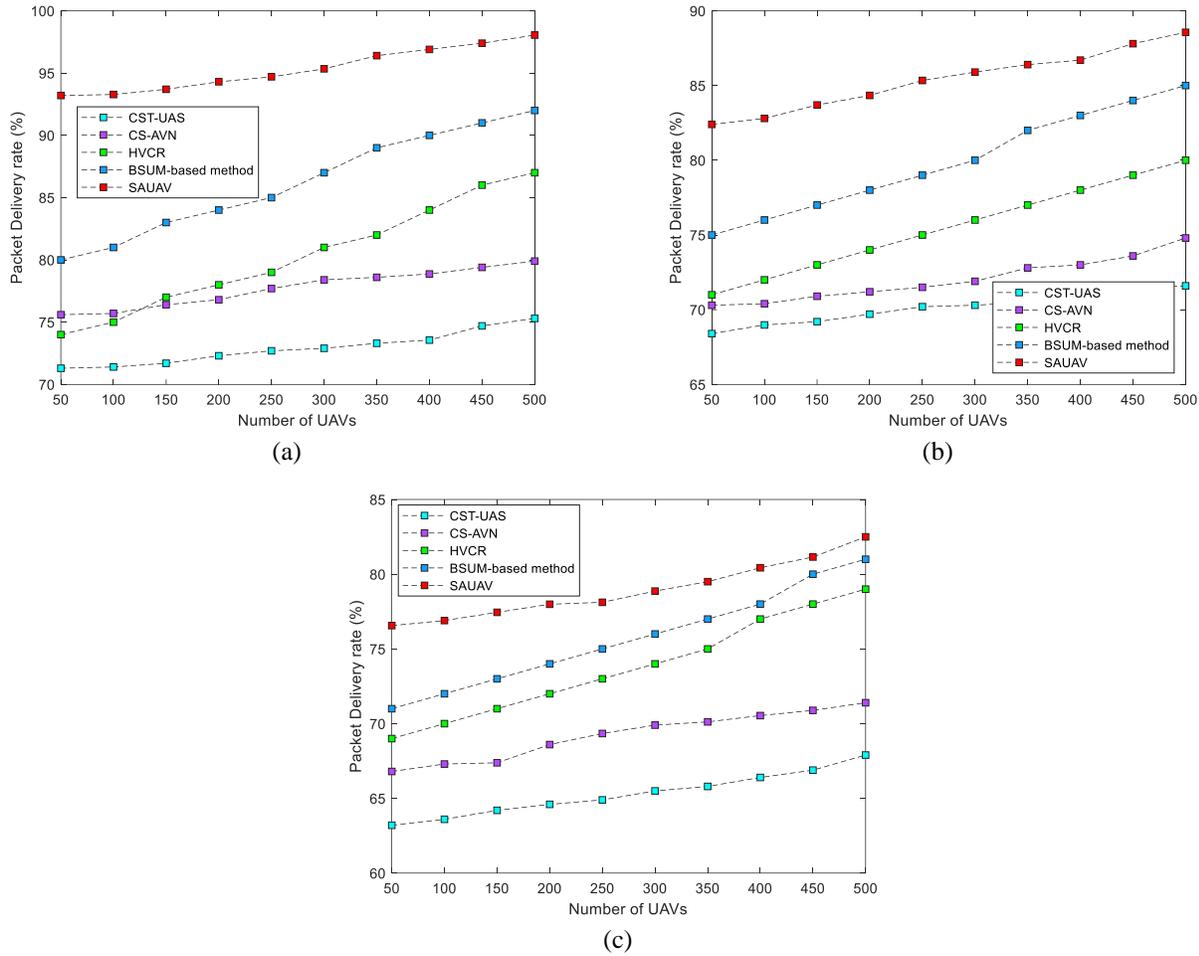

**Fig. 9** Comparison of the SAUAV, BSUM-based method, HVCR, CS-AVN, and CST-UAS models in term of *PDR*.

**RE:** Each UAV consumes energy to exchange packets between other UAVs. If the UAVs run out of energy, they will not be able to participate in the sending / receiving operation. Therefore, energy standard is a very important criterion for UAV networks. Because in the proposed method to secure the communication between the UAVs against Blackhole, sinkhole and Gray hole attacks, two phases based on the hash function are used as follows. In the first phase, in order to prevent the sending of fake UAV information to other normal UAVs, a series of safety rules were used to identify malicious operations to identify malicious UAVs and exclude them from routing operations. In the second phase, a mobile agent based on a three-step negotiation process was used to prevent the normal UAV from receiving false information from other malicious UAVs. Therefore, there will be no destructive UAVs in the network that waste the energy of other UAVs in a fake way. As a result, the amount of energy remaining in the proposed method will be higher than other methods. Therefore, while the malicious UAV rate is 10%, the amount of energy remaining in the SAUAV, BSUM-base, HVCR, CS-AVN and CST-UAS methods is 95%, 79%, 76%, 72% and 70%, respectively. This output for the second scenario is 91, 76, 73, 68 and 65, respectively. And for the third scenario, it is equal to 85, 74, 69, 66 and 62, respectively.

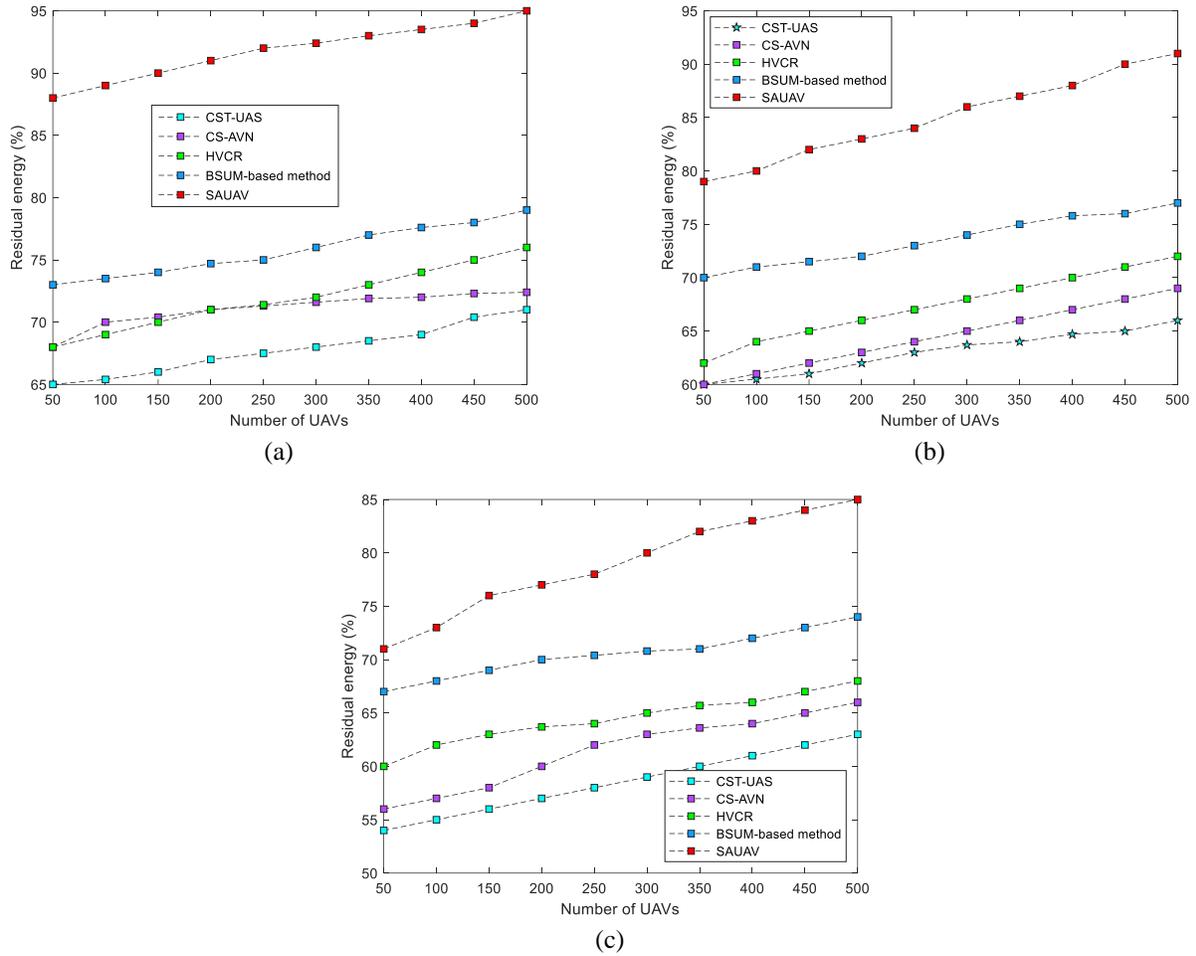

**Fig. 10** Comparison of the SAUAV, BSUM-based method, HVCR, CS-AVN, and CST-UAS models in term of *Residual energy*.

## 6 Conclusion and Future work

In this paper, two phases were used to secure communication between the UAVs. In the first phase, due to the behavior of malicious UAVs, in order to prevent sending fake information to other normal UAVs, a series of safety rules were used to identify malicious operations to identify malicious UAVs and prevent routing and data transfer operations between other UAVs. In the second phase, a mobile agent based on a three-step negotiation process was used to identify the mobile agent of each UAV from reliable neighbors through a three-step negotiation process, to prevent the normal UAV from receiving false information from other malicious UAVs. The combination of these two phases led to secure communications between the UAVs so that packets could exchange information securely. The simulation results showed that the proposed method was superior to CST-UAS, CS-AVN, HVCR, and BSUM-based methods in DR, FP, FN, PFR, and RE. In future work in this paper, a combination of two or more new and optimal algorithms such as earthworm optimization algorithm, moth search algorithm, monarch butterfly optimization, and elephant herding optimization will be used to identify malicious UAVs.

# Conflict of Interest

None.